\documentclass[epj]{svjour}
\usepackage{graphics}
\usepackage[sort&compress,numbers]{natbib}
\usepackage{doi}
\usepackage{hyperref}
\usepackage{amssymb,amsmath}
\usepackage{siunitx}
\usepackage{graphicx}
\usepackage[version=4]{mhchem}
\usepackage{subcaption}

\begin{document}
\title{Study of nuclear properties with muonic atoms}
\author{A.~Knecht\inst{1}\thanks{E-mail: a.knecht@psi.ch} \and A.~Skawran\inst{1,2}\thanks{E-mail: alexander.skawran@psi.ch} \and S.~M.~Vogiatzi\inst{1,2}\thanks{E-mail: stella.vogiatzi@psi.ch}
}                     
\institute{Paul Scherrer Institut, Villigen, Switzerland \and Institut f\"ur Teilchen- und Astrophysik, ETH Z\"urich, Switzerland}
\date{Received: date / Revised version: date}
%
\abstract{
Muons are a fascinating probe to study nuclear properties. Muonic atoms can easily be formed by stopping negative muons inside a material. The muon is subsequently captured by the nucleus and, due to its much higher mass compared to the electron, orbits the nucleus at very small distances. During this atomic capture process the muon emits characteristic X-rays during its cascade down to the ground state. The energies of these X-rays reveal the muonic energy level scheme, from which properties like the nuclear charge radius or its quadrupole moment can be extracted.\\
While almost all stable elements have been examined using muons, probing highly radioactive atoms has so far not been possible. The muX experiment has developed a technique based on transfer reaction inside a high pressure hydrogen/deuterium gas cell to examine targets available only in microgram quantities.
\PACS{
      {14.60. Ef}{Muons}   \and
      {36.10.Dr}{Positronium, muonium, muonic atoms and molecules} \and
      {32.30.Rj}{X-ray spectra} \and
      {21.10.-k}{General and average properties of nuclei}
     } 
} 
\maketitle
\section{Introduction} \label{sec_intro}
Muons are fascinating particles with experiments being conducted in the context of particle, nuclear and atomic physics \cite{gor15}. Additionally, also applied research is possible by measuring the spin precession and dynamics of muons inside materials through the $\mu$SR technique \cite{blu99} thus probing the internal magnetic fields of the sample under study.

Muons are so-called leptons -- such as electrons -- and classified into the second family within the Standard Model of Particle Physics \cite{tan18}. The properties of the muon have been extensively studied over the past 84 years since its discovery in 1936 \cite{and36, ned37}. Its lifetime \cite{tis13}, e.g., is the best measured lifetime of any unstable particle and searches for rare decays of the muon violating lepton family number have reached unprecedented sensitivity \cite{ber13}. With a mass of 105.66~MeV/c$^2$ they are about 207 times heavier than an electron. Their lifetime of 2.197~$\mu$s is short, but still long enough in order to allow the large range of experiments alluded to above. Muons are also at the center of several current discrepancies between the Standard Model of particle physics and experimental results: i) the discrepancy between the measured anomalous magnetic moment and its prediction \cite{mil12}, ii) discrepancies in the rare decays of $B$ mesons involving muons \cite{bla17}, and iii) the discrepancy of the measured proton radius between the experiments using a muon as a probe compared to the ones involving electronic probes \cite{poh10}. It is thus currently an exciting time to pursue experiments involving muons and who knows what new secrets these fascinating particles will reveal.

Muons are constantly being produced in the the upper atmosphere through the interaction of cosmic rays with the remaining air. Due to their high energy, a large fraction of them reaches all the way to the ground and even penetrates deep into the rock \cite{all85, cro78}. As such these cosmic muons are part of the natural background. However, with a flux at sea level of around 1~muon/cm$^2$/minute the intensity is too low to perform precision experiments. For this reason, the interactions happening in the upper atmosphere are replicated through the bombardment of a target (typically a low-Z material such as graphite) with a proton beam from an accelerator. The muons are not produced directly, but follow from the decays of pions created in the following interactions of the primary proton with the target nucleons: 
\begin{equation}
\begin{tabular}{lcl}
p + p $\rightarrow$ p + n + $\pi^+$ & & p + n $\rightarrow$ p + n + $\pi^0$ \\
p + p $\rightarrow$ p + p + $\pi^0$ & & p + n $\rightarrow$ p + p + $\pi^-$ \\
p + p $\rightarrow$ d + $\pi^+$ & & p + n $\rightarrow$ n + n + $\pi^+$  \\
& & p + n $\rightarrow$ d + $\pi^0$ 
\end{tabular}
\end{equation}
These are only the single pion production channels. For higher proton energies higher multiplicity pion production is possible.

\begin{figure}[]
\begin{centering} 
\resizebox{0.9\textwidth}{!}{
\includegraphics{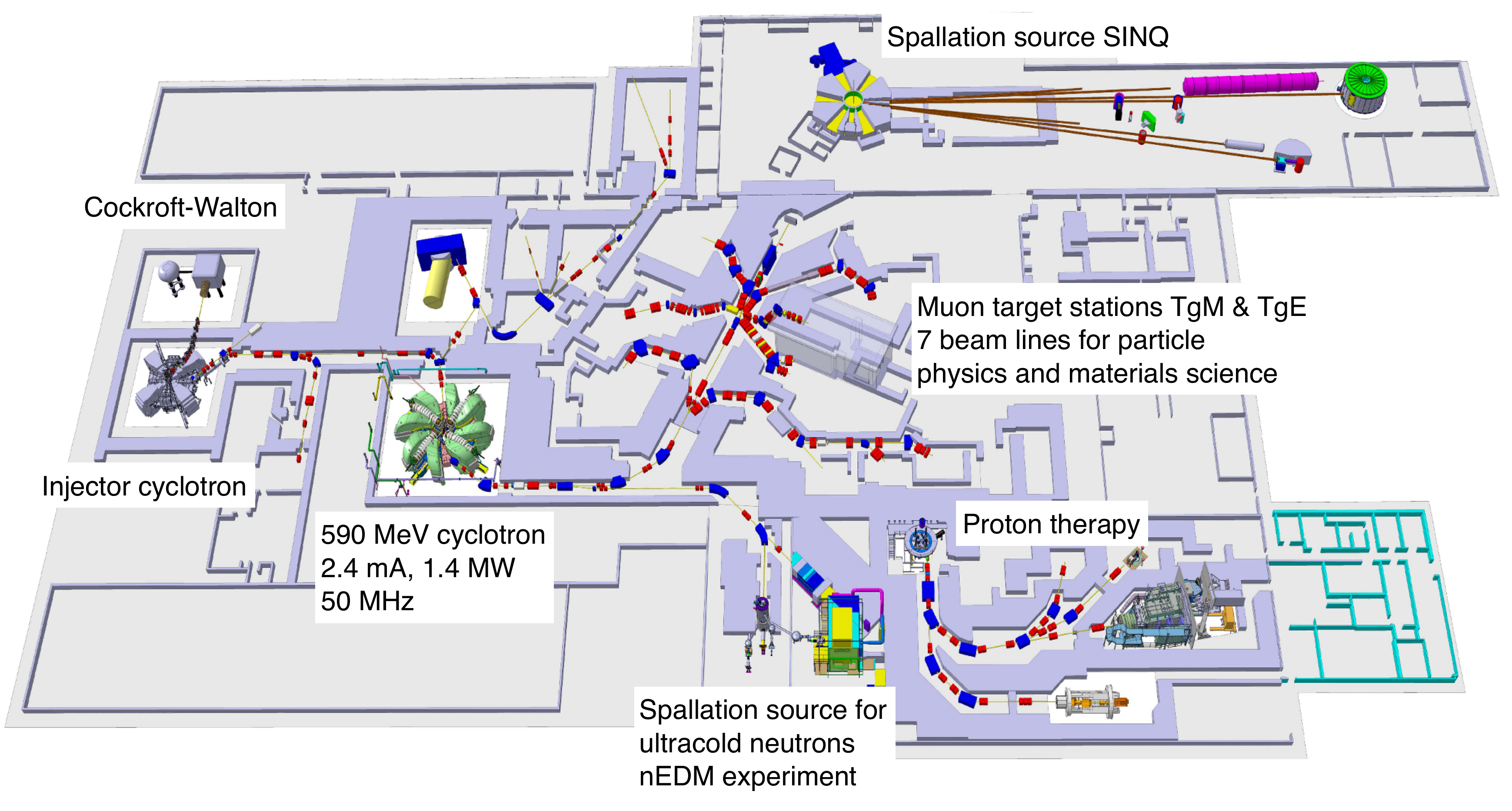}
}
\caption{Picture showing the proton accelerator complex HIPA at the Paul Scherrer Institute \cite{psi}. The proton accelerator serves two spallation targets for neutron production and two target stations for the production of muons.}
\label{fig_hipa}   
\end{centering}
\end{figure}

Figure~\ref{fig_hipa} shows the HIPA proton accelerator complex at the Paul Scherrer Institute (PSI) in Switzerland \cite{psi} where the muX experiment described in Section~\ref{sec_muX} takes place. Its 590~MeV proton beam reaches a world-leading power of 1.4~MW. Two target stations are available for muon production serving a total of 7 beam lines for particle physics and materials science. The highest intensities available at a low momentum of 28~MeV/c (very suitable for precision physics as these muons can easily be stopped) are $5\times 10^8$~$\mu^+$/s and $7\times 10^6$~$\mu^-$/s \cite{pro08} many orders of magnitude more than what is available from cosmic rays.

Muonic atoms are atoms where a negative muon has been captured by the nucleus. Due to its 207 times higher mass compared to the electrons, the muon orbits the nucleus at a 207 times closer distance and experiences 207 times higher binding energies (neglecting in both cases the finite size of the nucleus). Some of the electrons are usually kicked out during the capture process, but are often very quickly refilled and are generally not perturbed by the presence of the muon (and vice versa). However, the negative muon effectively shields one of the protons such that for the electrons the nucleus of charge $Z$ appears like a nucleus of charge $Z-1$. The details about the nucleus that can be extracted by this atomic capture and subsequent cascade down to the muonic ground state through a method called muonic atom spectroscopy is the topic of the next Section~\ref{sec_xrays}. The muX experiment that was built to exploit this method and extend it to target materials that are only available in microgram quantities (such as highly radioactive targets) is described in Section~\ref{sec_muX}.

\begin{figure}[]
\begin{centering} 
\resizebox{0.4\textwidth}{!}{
\includegraphics{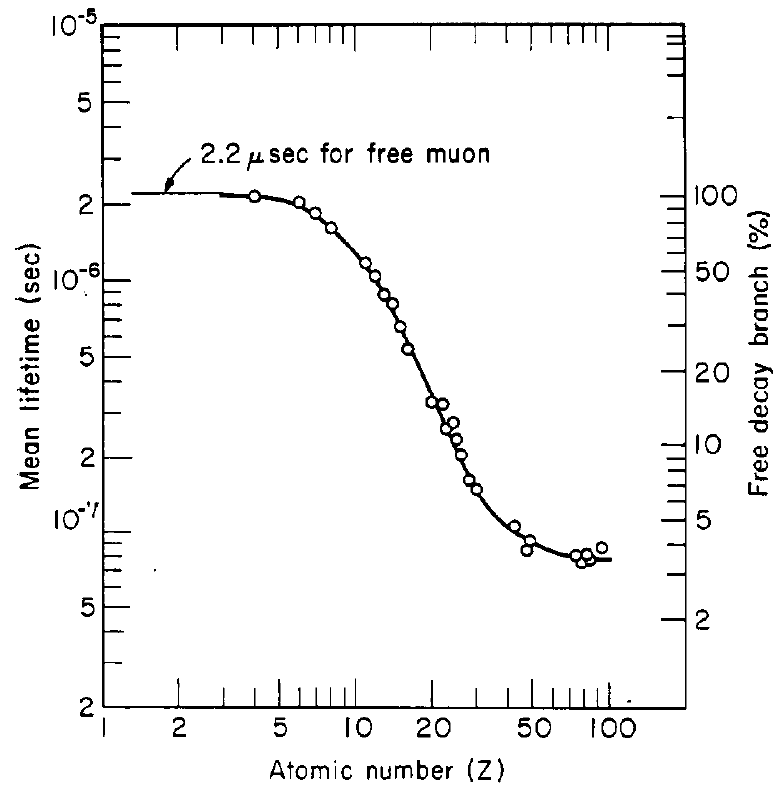}
}
\caption{Graph showing the mean lifetime of negative muons inside a material of atomic number $Z$ and the corresponding percentage of remaining ordinary muon decays. Reproduced from \cite{yam75}.}
\label{fig_lifetime}   
\end{centering}
\end{figure}

This paper only describes the knowledge that can be obtained on the nucleus through the study of the muonic X-rays emitted in the process of the atomic capture. However, much more nuclear physics is in principle involved in the whole interaction of the negative muon with the nucleus over its complete lifetime \cite{mea01}. Once the muon reaches the ground state, the overlap of the muonic wavefunction with the nucleus is so large -- as a matter of fact for high-Z elements the muon spends most of its time within the nucleus -- that the nuclear capture channel Eq.~(\ref{eq_capture}) becomes dominant over the ordinary decay of the muon Eq.~(\ref{eq_decay}).
\begin{eqnarray} 
\mu^- & \longrightarrow & e^-  \nu_\mu \bar{\nu}_e  \label{eq_decay} \\ 
\mu^- + p & \longrightarrow & n +  \nu_\mu  \label{eq_capture}
\end{eqnarray}
As shown in Fig.~\ref{fig_lifetime}, the total lifetime for negative muons in medium and high-Z elements is reduced to below 100~ns with only a few percent of the muons still undergoing ordinary muon decay. After this nuclear capture, the resulting daughter nucleus of charge $Z-1$ can be  highly excited and most frequently returns to its ground state through the emission of neutrons and characteristic gamma rays. The study of these gamma rays can reveal details of the nuclear structure of the daughter nucleus -- see, e.g., Ref.~\cite{zin18} where this technique has been employed to study nuclei relevant for neutrinoless double-beta decay.

\section{Muonic atom spectroscopy} \label{sec_xrays}
A muonic atom is an atom in which a negatively charged muon is captured by a nucleus. To form such an atom, a target material is bombarded with a negative  muon beam. The muons suffer successive kinetic energy losses by their interaction with the outer atomic electrons in the target. When the muon is sufficiently slowed down to have similar velocity as the atomic electrons, it is attracted by the Coulomb field of the nucleus and forms a muonic atom. The capture is followed by the cascade of the muon to the ground state of the atom with the emission of Auger electrons and X-rays. Since the energy of the muonic X-rays is determined by the exact details of the interaction of the muon with the nucleus, muonic atom spectroscopy is a sensitive tool to probe the properties of the nucleus.

\subsection{Muonic cascade}
\begin{figure}
\centering
\resizebox{0.6\textwidth}{!}{\includegraphics{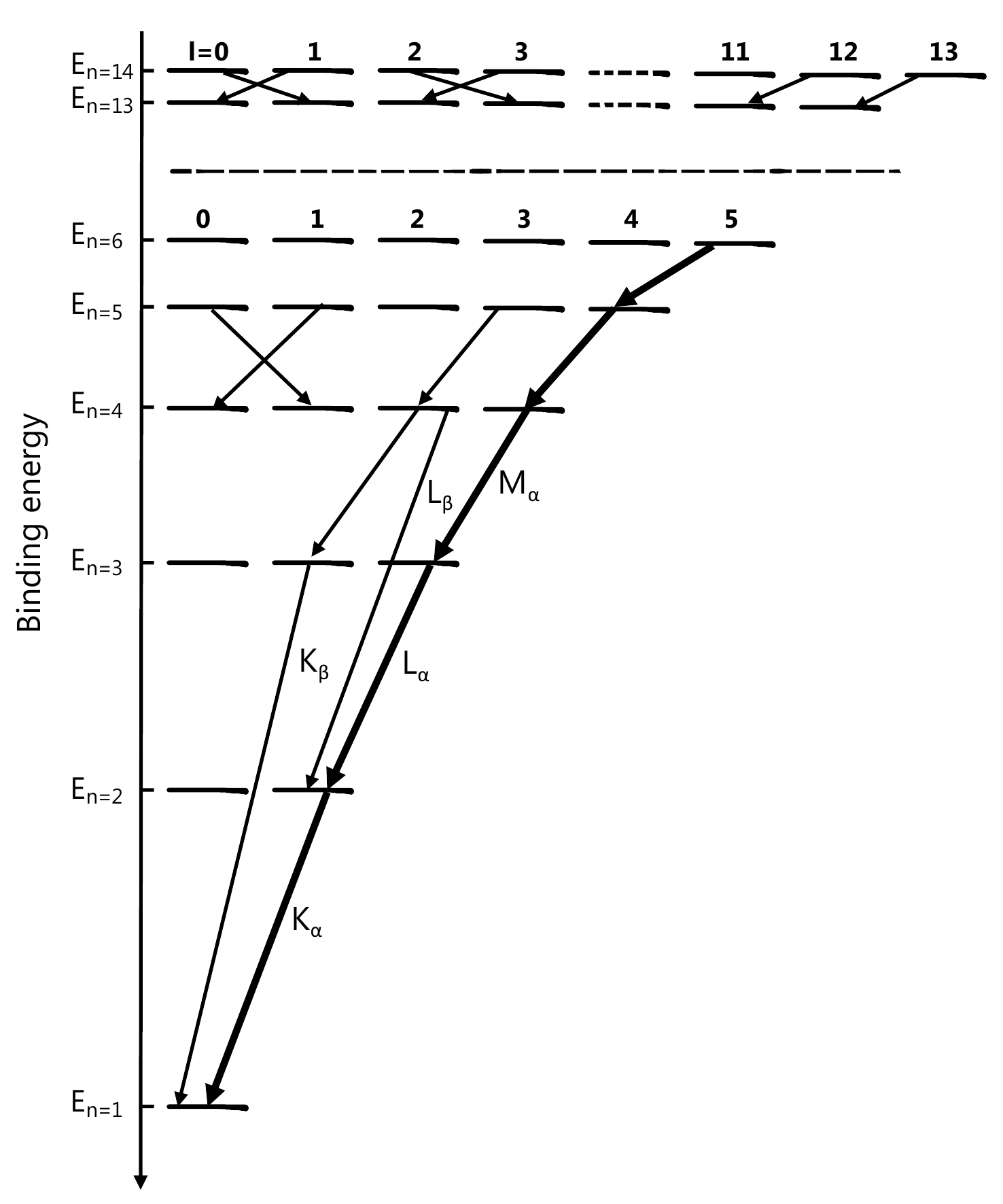}}
\caption{Simplified scheme of the muonic atom levels for a point-like nucleus. The arrows indicate the transitions of the muon through the different atomic levels.}
\label{fig_muonic_transitions}
\end{figure}

\begin{figure}
\centering
\resizebox{0.75\textwidth}{!}{\includegraphics{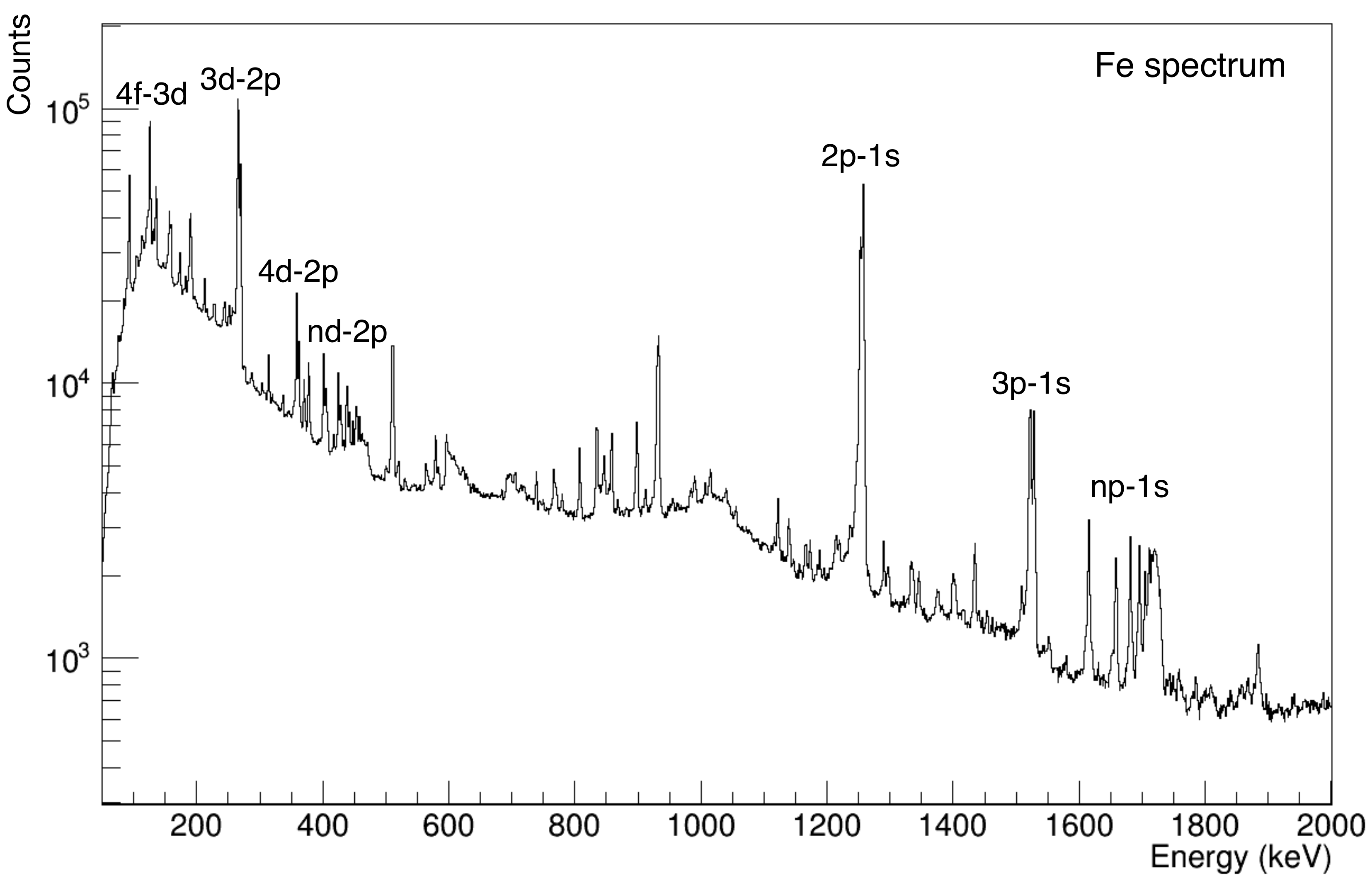}}
\caption{Full-energy spectrum in natural iron with the $K$, $L$ and $M_{\alpha}$ X-rays labelled.}
\label{fig_Fe_Xrays}
\end{figure}

The muon is captured at a high principal quantum number $n\sim 15$. The initial population of the levels is not well known but believed to approximately follow the $\sim (2l+1)$ distribution (with $l$ being the quantum number of the orbital angular momentum). From this short-lived state, the muon starts its cascade down to the ground state ($1s$) of the atom. The energy of the muonic transitions at the beginning of the cascade, where the muonic levels are relatively dense, are small but still much larger than the typical binding energies of the electrons. Hence, the cascade of the muon initially proceeds through the release of Auger electrons. For lower transitions below $n\simeq 6$ (where the energy gaps are much larger), the emission of muonic X-rays become the dominant de-excitation mechanism. These X-rays are characteristic for each element and their energy varies from a few keV to a few MeV. A simplified scheme of the muon cascade is illustrated in Fig.~\ref{fig_muonic_transitions}. The most intense transitions are the $(n,l=n-1)\rightarrow (n-1,l=n-2)$ transitions such as the $4f\rightarrow 3d$, $3d\rightarrow 2p$, $2p\rightarrow 1s$ etc. Figure~\ref{fig_Fe_Xrays} shows the energy spectrum of iron as measured with germanium detectors illustrating the muonic X-ray transitions. The whole cascade is very fast taking typically much less than a nanosecond. As described above, in the ground state of the muonic atom, the muon will disappear either via the free muon decay with a lifetime of \SI{2.2e-06}{\second} or through the nuclear capture process. 

\subsection{Muonic level scheme}
As only one muon orbits the nucleus at the same time and the remaining electrons are far away from the muon orbits, the muonic atom is essentially a hydrogen-like atom and the calculation of the muonic energy levels follows the steps found in many textbooks on the hydrogen atom \citep{dem06, hak05, eid07}. The gross features of the muonic energy levels are obtained in the Bohr approximation of a point-like nucleus. This can be done by solving the Schr\"{o}dinger equation assuming the electrostatic interaction between the muon and the nucleus. A more accurate description of the energy levels is given by the Dirac equation. The Dirac equation in natural units for a muon orbiting in the field of the nucleus is expressed as:
\begin{equation}
(\vec{\alpha}\cdot\vec{p}_{\mu} + \beta \, m_{\mu})\,\psi(\vec{r}) = (E-V(\vec{r}))\,\psi(\vec{r})
\label{eq_Dirac}
\end{equation}
where $\vec{\alpha}$ and $\beta$ are the Dirac matrices, $\vec{p}_{\mu}$ the momentum of the muon, $\psi(\vec{r})$ the wave function of the bound muon at position $\vec{r}$ and $V(\vec{r})$ is the electrostatic potential generated by the nuclear charge distribution. A general expression of the nuclear potential is given by \cite{Bor82}
\begin{equation}
V(\vec{r}) = -Z \alpha \int d^3\vec{r}_N \frac{\rho(\vec{r}_N)}{\vert \vec{r}-\vec{r}_N \vert}
\label{eq_potential}
\end{equation}
where $\alpha$ is the fine-structure constant, $Z$ the nuclear charge and $\rho(\vec{r}_N)$ denotes the static charge distribution of the nucleus.

For a point-like nucleus the electrostatic potential takes the form $V_C(r)=-Ze^2r^{-1}$ and the resulting energy of the muonic levels is the sum of the Bohr energy and the fine-structure shift. The fine structure originates from the coupling of the spin $s=1/2$ with the orbital angular momentum of the muon $l$. The resulting quantum number of this $ls$-interaction is the total angular momentum of the muon $j$. Taking the values $\vert l \pm 1/2 \vert$, the new quantum number $j$ lifts the degeneracy of the muonic energy states such that the $2p$ level splits into the $2p_{1/2}$ and $2p_{3/2}$ states (following the notation $nl_j$).

\begin{figure}
\centering
\resizebox{0.55\textwidth}{!}{\includegraphics{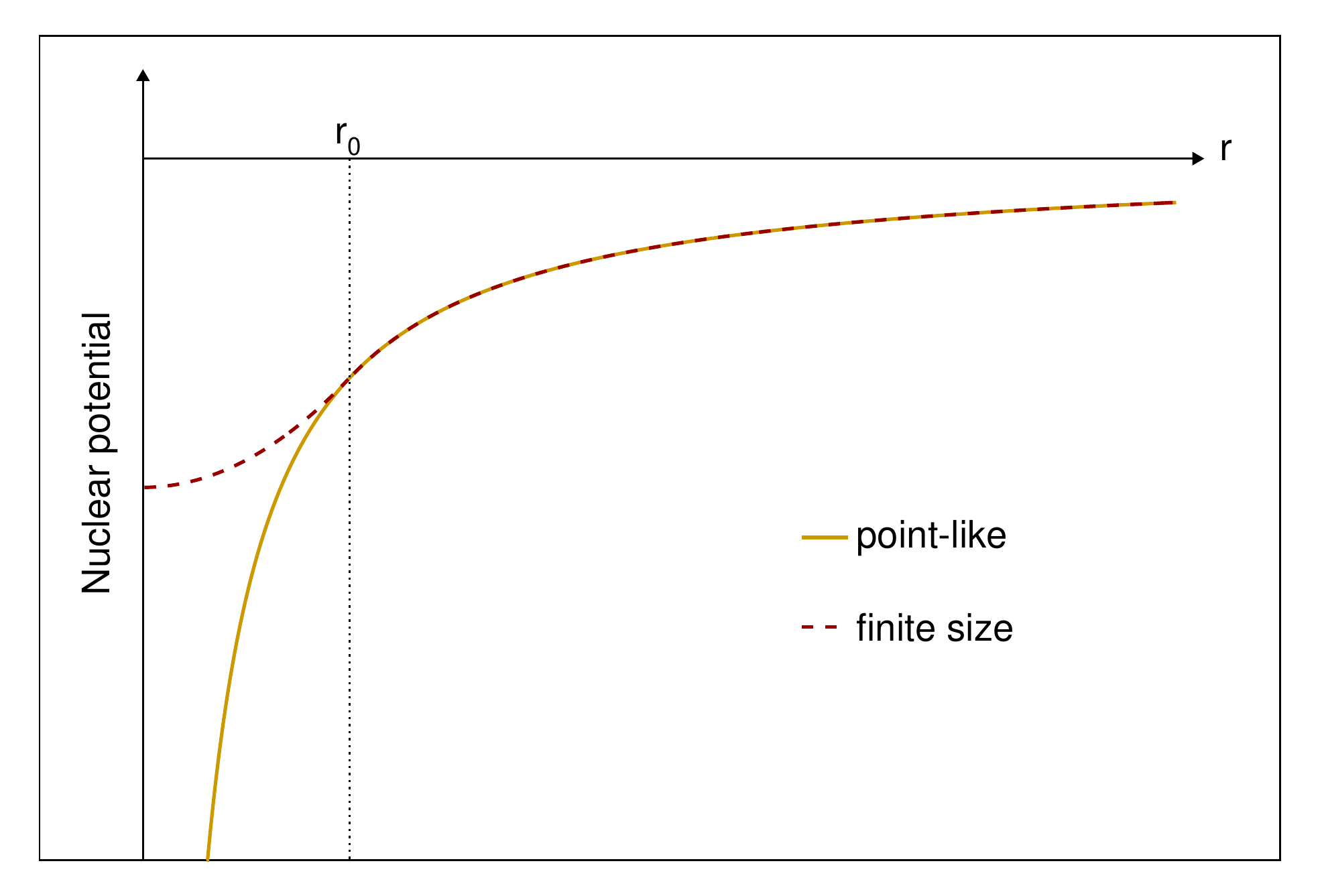}}
\caption{The nuclear potential as a function of the distance from the center of the nucleus $r$ for a point-like nucleus (solid line) and for a nucleus approximated by a homogeneously charged sphere of radius $r_0$ (dashed line).}
\label{fig_nuclearPotential}
\end{figure}

For the low-lying levels the point-like nucleus is a very strong approximation which only very roughly describes the muonic energy levels. The effect due to the finite size of the nucleus is typically calculated by numerically solving the relativistic Dirac equation including the nuclear potential under the assumption of some reasonable nuclear charge density. Such a commonly used charge distribution for spherical nuclei is the two parameter Fermi distribution given as 
\begin{equation}
\rho(r)= \rho_0 \left\lbrace 1+\exp\left( 4\, \ln3\,\frac{r-c}{t} \right) \right\rbrace ^{-1}
\label{eq_FermiSpherical}
\end{equation}
where $t$ is the surface thickness defined as the distance at which the charge density reduces from 90\% to 10\% of the density at the center of the nucleus $\rho_0$ and $c$ describes the half-density radius. Due to the mass of the muon which is $\sim 207$ times larger than the mass of the electron, the binding of the muon to the nucleus is significantly stronger than the binding of the electron. Therefore, the low-lying muonic states hugely overlap with the nuclear charge distribution. For $n\gtrsim 6$ the muon is sufficiently far away from the nucleus such that the finite size corrections are very small. In the presence of a finite nuclear density, the nuclear potential changes drastically as shown in Fig.~\ref{fig_nuclearPotential}. This causes a reduction of up to \SI{50}{\percent} of the binding energy of the low-lying states in heavier nuclei with respect to the point-like nucleus approximation \cite{Sal17}.

For nuclei which suffer a shape distortion, the nuclear charge distribution is developed as a multipole expansion $\rho(\vec{r}) = \rho_0(r) + \rho_2(r)\,Y_{20}(\theta)$ where $\rho_0(r)$ is the spherically symmetric part that determines the gross and fine structure while the second term is the non-spherical part relevant for the hyperfine splitting described below. The Fermi distribution then contains two extra parameters ($\beta$, $\gamma$) that define the shape deformation of the nucleus as \cite{ack66}
\begin{equation}
\rho(\vec{r},c,t) = \rho_0 \left[ 1+\exp \left( 4\, \ln3 \, \frac{r-c \left( 1+\beta \, Y_{20} \right)}{ t \, (1+\beta\,\gamma Y_{20})}   \right) \right] ^{-1} \,.
\label{eq_FermiDeformed}
\end{equation}

The muonic energy levels experience an additional splitting, which is known as the hyperfine splitting (HFS). This effect arises from the coupling of the total angular momentum of the muon $J$\footnote{While the total angular momentum is usually written as $j$ for the fine splitting, it is given as $J$ in the hyperfine splitting.} with the nuclear spin $I$ and is described by a new quantum number $\vec{F}=\vec{I}+\vec{J}$. The HFS effect is typically smaller than the fine splitting except for highly deformed nuclei where they become of the same order. The allowed transitions of the muon are given according to the selection rules of this new quantum number $F$. The HFS occurs either in deformed nuclei whose shape is not spherical or in nuclei with non-zero magnetic moment and involve the electric quadrupole (E2) and magnetic dipole (M1) interactions. The energy shift of the muonic state of total angular momentum $F$ due to the M1 and E2 interaction accounting for the distribution of the nuclear moments is given by the generalized Bohr-Weisskopf formulas as \cite{leb63}
\begin{equation}
\Delta E(M1) = \frac{A_1}{2}\left\lbrace F(F+1)-I(I+1)-J(J+1)  \right\rbrace
\label{eq_M1}
\end{equation}
\begin{equation}
\Delta E(E2) = A_2 \frac{6 \left[ K(K+1)-\frac{4}{3} I(I+1)J(J+1)\right]}{4I(2I-1)J(2J-1)}
\label{eq_E2}
\end{equation}
where $K=F(F+1)-I(I+1)-J(J+1)$. The M1 hyperfine constant $A_1$ is determined by the magnetic interaction between the muon and the nucleus in terms of the spin and orbital magnetization. It depends on the spatial extent of the nuclear magnetic moment and the wave function and the magnetic moment of the orbiting particle. The HFS constant $A_2$ is a function of the nuclear quadrupole moment generated by the non-spherical part of the nuclear charge distribution. The quadrupole moment describes the effective shape of the nucleus taking a value of zero for spherically symmetric, positive values for prolate and negative values for oblate nuclei. Due to the muon to electron mass ratio, the magnetic moment of the muon is $\sim 207$ times smaller than the one of the electron. In muonic atoms, therefore, the quadrupole deformation is two orders of magnitude stronger than the magnetic effect contrary to the electronic atoms. However, the magnetic part of the HFS can be studied in spherical nuclei where the quadrupole interaction is eliminated.

In atoms where the muon is bound to a highly deformed nucleus, the energy of several of the muonic transitions is of the same order as the excitation energy of the nucleus to states with non-zero spin. Due to the $IJ$-coupling, this leads to an admixture of ground and excited hyperfine states and a more complicated structure of the muonic energy levels. The effect is called dynamic E2 interaction and is  responsible for the presence of the hyperfine splitting in even-even nuclei \cite{hit70, mic19, Dev69}. 

\begin{figure}
\centering
\resizebox{0.55\textwidth}{!}{\includegraphics{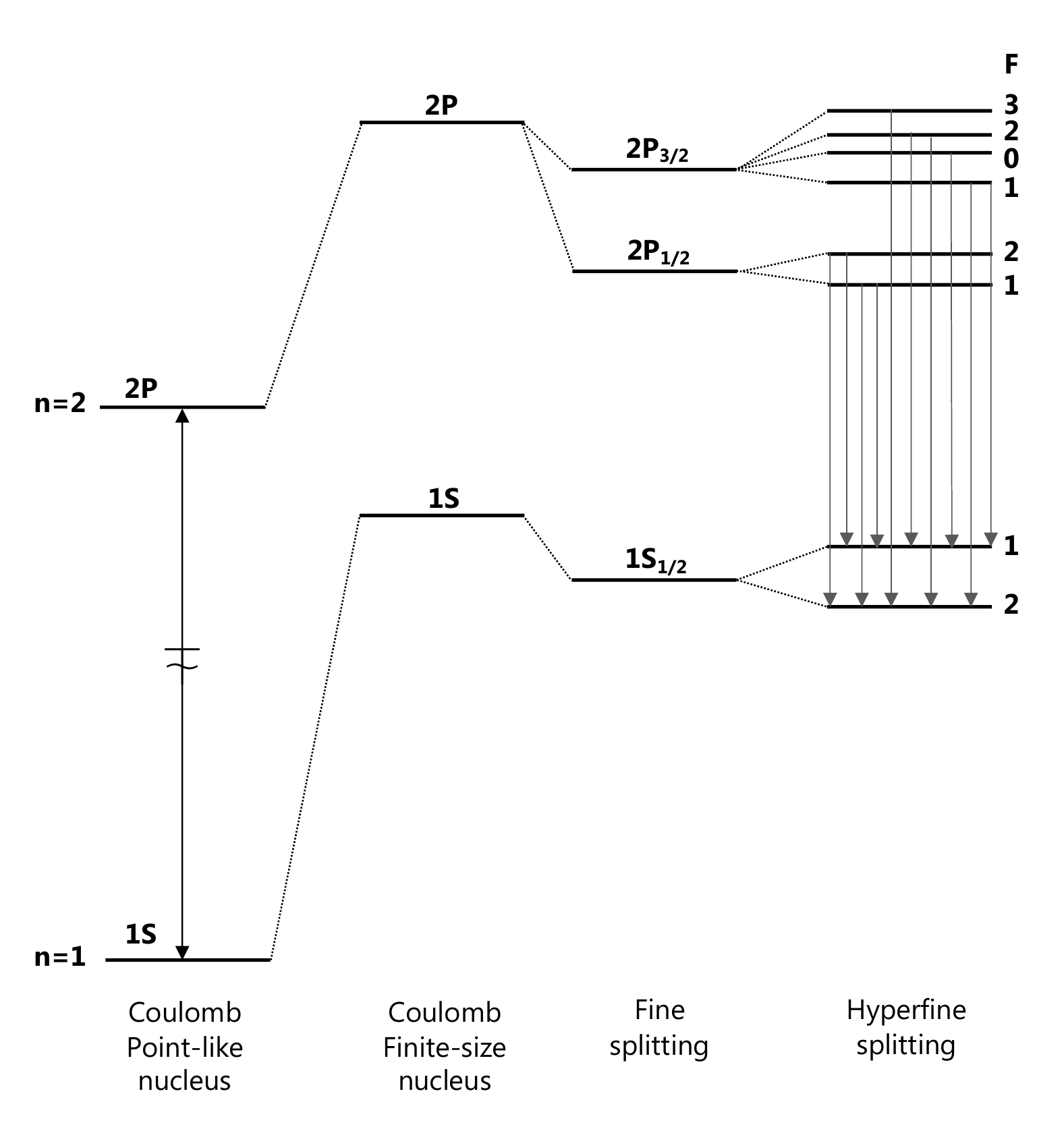}}
\caption{Schematic (not to scale) of the muonic energy levels for a nucleus with spin $I=\frac{3}{2}$ in the ground state showing the effect of the finite nuclear size, fine and hyperfine splitting. The arrows illustrate the allowed hyperfine transitions of the muon according to the selection rules $\Delta F=0,\pm1$.}
\label{fig_muonicLevels}
\end{figure}

A schematic of the finite size, fine and hyperfine splitting is shown in Fig.~\ref{fig_muonicLevels}. Other corrections to the muonic levels are the vacuum polarization, the Lamb shift, the nuclear polarization, the electron screening and the anomalous magnetic moment. More details about the theory of muonic atoms can be found in, e.g., \cite{Dev69, Fri04, Kim71, mea01, mck69, dew66, dey79, wu69}.

\subsection{Extraction of charge radii}
Profiting from the high sensitivity of the low-lying muonic transitions to the properties of the nuclear charge distribution, the charge radii of almost all stable elements have been determined employing muonic atom spectroscopy. In order to do so, the energy of the muonic levels is  calculated accounting for all the relevant effects and given as a function of the charge distribution parameters. The parameters of the charge distribution can subsequently be extracted when the experimentally observed transitions are matched to the predicted function. However, typically it is the root-mean-square radius which is reported rather than the charge distribution parameters. The r.m.s. radius is given as $r_C = \sqrt{ \left\langle r^2 \right\rangle}$, where $\left\langle r^2 \right\rangle$ is the second radial moment defined as \cite{Fri04}
\begin{equation}
\left\langle r^2 \right\rangle = \frac{1}{Ze} \int d^3\vec{r} \rho(\vec{r}) \, r^2 \,.
\end{equation}

The most difficult component that needs to be treated in the theoretical analysis and in the end limits the uncertainty on the charge radius extraction is the nuclear polarization correction. This effect emerges due to the electromagnetic interaction between the muon and the nucleus exciting the system into virtual states. The effect results in a shift of the muon binding energies which for the 1s state in \ce{^208Pb} is \SI{6}{\kilo\electronvolt} \cite{che70a}. Due to inadequate knowledge of the nuclear excited states, the uncertainty of the calculation of the nuclear polarization limits the experimental accuracy. Despite this limitation, very precise values for the charge radius can be obtained. For $^{208}$Pb, e.g., the charge radius was determined with a relative precision of \SI{0.02}{\percent} \cite{Ber88}.

\subsection{Extraction of quadrupole moments}\label{sec_quadmoments}
The extraction of quadrupole moments from measured muonic X-rays spectra has been performed many times in the past - as examples see Refs.~\cite{dey79, kon81, del87, pow76}. Here, we briefly describe our extraction of the quadrupole moments for isotopically pure \ce{^{185,187}Re} targets \cite{mich19,ant20}. 

In heavy muonic atoms there exists an intermediate domain of energy states such as $n=5$, $n=4$ where the radius of the muon orbits is larger than the nuclear size and the muon is not influenced by the presence of the surrounding atomic electrons. In this regime, the hyperfine constants $A_1$ and $A_2$ (shown in Eq.~(\ref{eq_M1}) and (\ref{eq_E2})) are independent of the details of the electric quadrupole and magnetic moment distributions. Therefore, the measurement of the quadrupole energy splitting is a direct probe of the spectroscopic quadrupole moment which can be obtained by fitting the experimentally observed $5g\rightarrow 4f$ hyperfine transitions.

The $5g_{9/2}\rightarrow 4f_{7/2}$ and $5g_{7/2}\rightarrow 4f_{5/2}$ hyperfine complexes in \ce{^{185,187}Re} are treated together with three weaker multiplets which coincide in energy, namely the $5f_{7/2}\rightarrow 4d_{5/2}$, $5g_{7/2}\rightarrow 4f_{7/2}$ and $5f_{5/2}\rightarrow 4d_{5/2}$. Due to the nuclear spin of the rhenium isotopes in the ground state of $I=5/2$, the hyperfine pattern consists of 76 transitions which appear as two bumps at around \SI{360}{\kilo\electronvolt}. To disentangle the hyperfine lines, a very good knowledge of the instrumental line-shape, which describes the detector's response, is required. The line-shape model was fitted to four background and nuclear capture lines close in energy to the hyperfine structure providing a consistent set of line-shape parameters. The obtained parameters were then fixed for each of the 76 transitions and a global fit of the $5\rightarrow 4$ hyperfine structure was performed. The intensities and energies of the hyperfine lines were expressed as a function of the quadrupole moment up to a quadratic term with regards to the $F=7\rightarrow F=6$ transition of the $5g_{9/2}\rightarrow 4f_{7/2}$ multiplet. In this way, the quadrupole moments could be treated as parameters of the global fitting function and were determined by minimizing the overall $\chi^2$ of the fit. 

The extracted nuclear spectroscopic quadrupole moment is \SI{2.07(5)}{\barn} and \SI{1.94(5)}{\barn}  for \ce{^{185}Re} and \ce{^{187}Re}, respectively. The results were compared with older literature values obtained by  means of muonic X-ray spectroscopy in a natural rhenium sample \cite{kon81}. A disagreement at the order of \SI{7}{\percent} was found. The discrepancy is attributed to the higher sensitivity of the isotopically pure data to the weaker transitions (neglected in the previous determination) involved in the fitting of the $5\rightarrow 4$ hyperfine structure.

\section{The muX experiment} \label{sec_muX}
In the following pages, we briefly describe the muX experiment which developed a new method to perform muonic atom spectroscopy with targets that are only available in microgram quantities. This enables the application of muonic atom spectroscopy to highly radioactive isotopes and elements that are only available in small quantities.

As described above, the measurement of muonic X-rays is performed by stopping a negative muon beam in a target resulting in the atomic capture of the muon and the subsequent cascade down to the ground state leading to the emission of the characteristic X-rays. A typical experimental setup contains an entrance detector providing the start signal for an incoming muon. The target is mounted behind this detector and surrounded by veto detectors in order to detect and veto the electrons from the ordinary muon decay. Typically, the entrance and the veto detectors are plastic scintillators. The detectors for the measurement of the muonic X-rays are arranged behind the veto detectors. The best choice in order to obtain good energy resolution are High-purity germanium (HPGe) detectors. Such a setup was assembled for the measurement of the stable elements $^{185}$Re and $^{187}$Re by the muX collaboration \cite{ant20} briefly mentioned in Section~\ref{sec_quadmoments}.

However, one of the main goals of the muX experiment is the measurement of the radioactive isotopes $^{226}$Ra and $^{248}$Cm, which are limited to an amount of only a few $\SI{}{\mu g}$ in the PSI HIPA facility. The standard approach of directly stopping a conventional muon beam in such small target masses is not possible and a novel method for forming the muonic atoms needed to be developed. The new approach is based on the experience gained during the research on muon catalysed fusion concerning the behavior of negative muons in hydrogen gases (see \cite{pet01} and references therein). It relies on repeated muon transfers in a hydrogen-deuterium gas mixture and is inspired by the achievements of Abott et al. \cite{Abb97} and Strasser et al. \cite{Str09}.

\begin{figure}
\centering
\includegraphics[width=0.5\textwidth, angle=90]{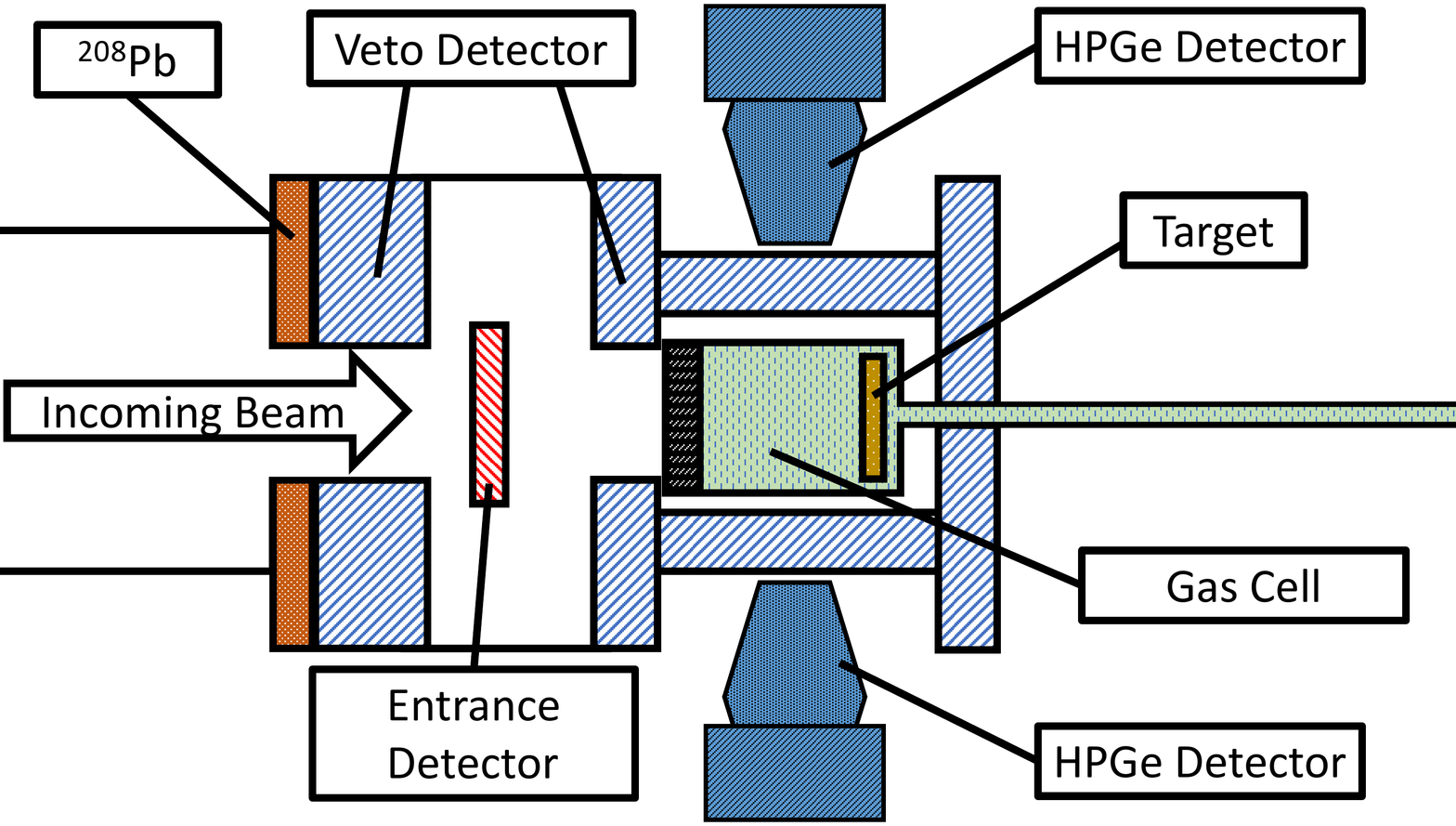}
\caption{Sketch of the newly developed gas cell setup for muonic atom spectroscopy experiments. The negative $\mu$ beam enters the setup from the left side. Incoming muons are registered by the entrance detector.  The muons, which enter the gas mixture, are stopped and transferred in three steps to the target mounted at the backside of the cell. The result is a muonic X-ray cascade in the target. For calibration purposes $^{208}$Pb is mounted in front of the entrance detector to constantly emit muonic X-rays of well-known energy. The muonic X-rays are detected by HPGe detectors arranged around the target. Plastic scintillators are used for the entrance and veto detectors.}
\label{Fig:ExpSetup}
\end{figure}

\subsection{Transfer reactions in a hydrogen/deuterium gas cell}
Figure~\ref{Fig:ExpSetup} shows a sketch of the experimental apparatus. While the details of the setup will be described in Section~\ref{sec_setup}, the sketch shows that at the center of the apparatus a gas cell is located with the target mounted within its volume. The gas cell is filled with 100~bar hydrogen and a 0.25\% admixture of deuterium. The following paragraphs describe the processes occurring in this gas cell that allows to perform muonic atom spectroscopy with only a few microgram of target material.

\paragraph{Muon capture and transfer}
The process of the muon transfers to the target starts with a muon entering the gas cell. Due to the high pressure in the gas the muon is quickly stopped, captured by a hydrogen molecule and a $\mu$-molecular complex is created:
\begin{equation}
\left( ab \mu e \right)^* .\label{Equ:Formation}
\end{equation}
 $a$ and $b$ are hydrogen isotopes and $e$ is an electron. The prepared gas mixture contains natural hydrogen and a small admixture of deuterium. Hence, $a$ or $b$ is most likely a proton ($p$) and we can rewrite Eq.~\ref{Equ:Formation} as
\begin{equation}
\left( pp \mu e \right)^* .
\end{equation}
The main dissociation modes \cite{kor96} of the $\mu$-molecular complex result in the production of muonic hydrogen $\mu p$ by either direct dissociation
\begin{equation}
\left( pp \mu e \right)^* \rightarrow  \left( p \mu  \right)^* + \left(  pe \right)
\end{equation}
or by electron emission and subsequent dissociation
\begin{equation}
\left( pp \mu e \right)^* \rightarrow  \left( p p\mu  \right)^* + e \rightarrow \left( \mu p  \right)^* + p + e \,.
\end{equation}

Following the formation, the muonic hydrogen de-excites to the ground state and is thermalized through multiple scatterings with the surrounding hydrogen molecules. Due to the large cross section for $\mu  p$-H$_2$ scattering, only a very small number of muonic hydrogen atoms can travel far enough to reach the target.

\begin{figure}
\centering
\includegraphics[width=0.7\textwidth]{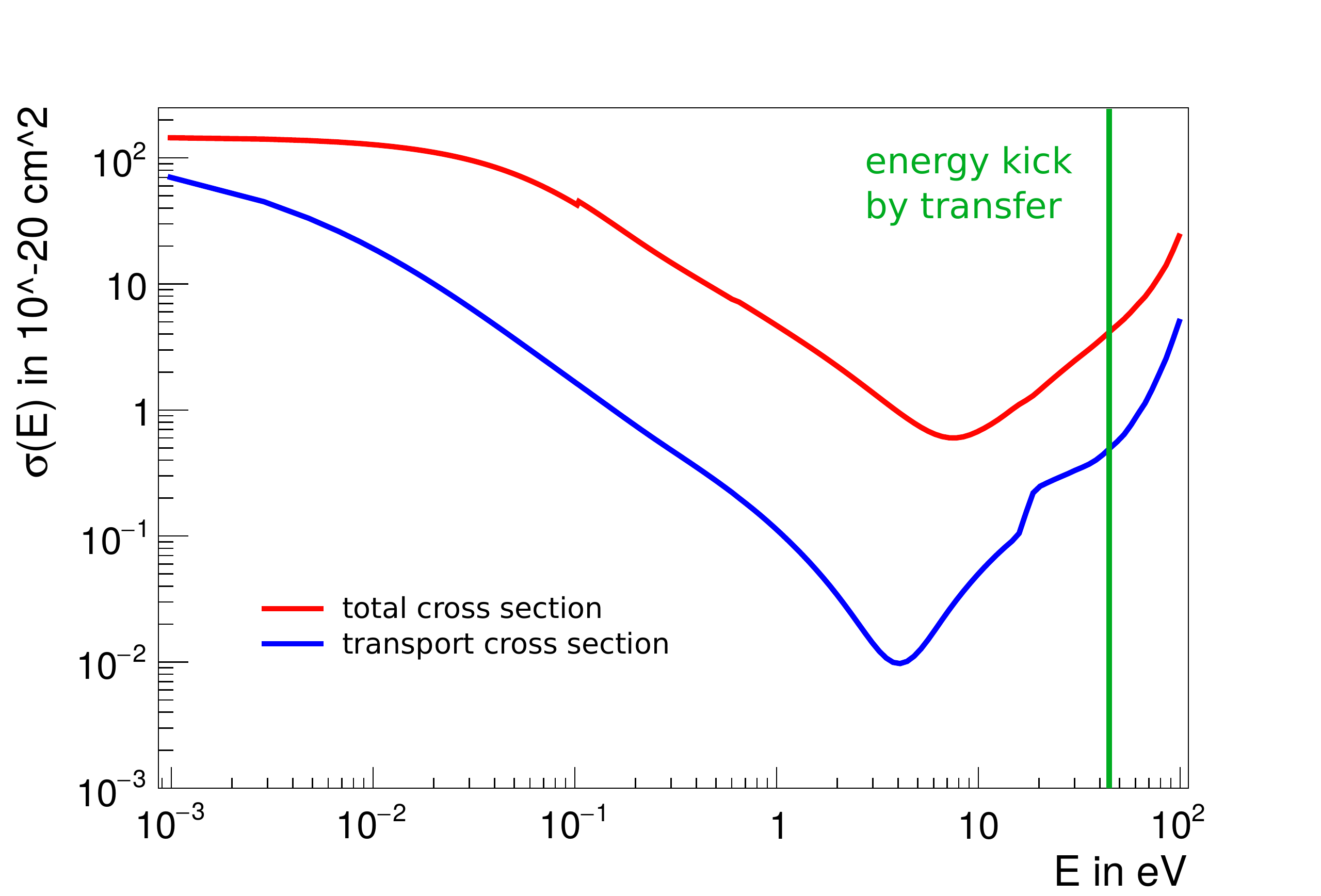}
\caption{Total and transport cross-sections for the elastic $\mu d$-H$_2$ scattering in the laboratory frame. The cross-sections feature a deep minimum at a kinetic energy of around $\SI{4}{eV}$ due to the Ramsauer-Townsend effect  \cite{Mul06,Ada96,ada07}. The kinetic energy gain due to the muon transfer from muonic hydrogen to muonic deuterium $\mu p + d \rightarrow p +\mu d$ is shown by the green line.}
\label{Fig:Ramsauer}
\end{figure}

The key to increase the distance over which the muonic atoms can travel lies in the so-called Ramsauer-Townsend effect \cite{ram21} which requires a small deuterium admixture in the hydrogen gas. When $\mu p$ scatters with a deuterium atom the muon is transferred from muonic hydrogen to muonic deuterium $\mu  d$\begin{equation}
\mu p + d \rightarrow p +\mu d^*,
\end{equation}
where $d$ is the deuteron. For the conditions used in the muX gas cell, the rate of this transfer amounts to $\sim 3.5 \times 10^6$/s \cite{and07}. The reason for this transfer is the larger binding energy of $\mu  d$ due to the higher mass of a  deuteron compared to a proton. From the released binding energy, the $\mu  d$ atom gains a kinetic energy of $\SI{45}{eV}$ \cite{Mul06} which is gradually lost by subsequent scatterings. 

At a kinetic energy of $\sim$$\SI{4}{eV}$~\cite{Mul06,Ada96,ada07}, the $\mu d$-H$_2$ scattering cross-section features a deep minimum due to the Ramsauer-Townsend effect (see Fig.~\ref{Fig:Ramsauer}) and the hydrogen basically becomes ``transparent'' to the $\mu d$ atom. The distance covered by the $\mu d$ atom over its lifetime can be best described by the mean diffusion radius $\overline{R}_{diff}$ \cite{ada07}. As the scattering cross-section is anisotropic, it is best to estimate $\overline{R}_{diff}$ with the so-called transport cross-section 
\begin{equation}
\sigma_{tran} = \int d\Omega (1-\cos \vartheta ) \frac{d \sigma(\vartheta)}{d \Omega},
\end{equation}
where $d\sigma ( \vartheta )/d \Omega$ is the differential cross-section of the $\mu d$-H$_2$ scattering process and $\vartheta$ is the scattering angle. $\sigma_{tran}$ is also termed momentum-transfer cross-section as the average momentum transfer $\Delta p = p_{in}-p_{out}$ in the scattering process is directly proportional to $\sigma_{tran}$. The calculation shows that $\overline{R}_{diff}\propto1/\sqrt{\sigma_{tran}}$ \cite{ada07}, which amounts to around 5~mm for the gas mixture used in the muX experiment (compared to the $\sim$0.5~mm for a thermalized $\mu p$ atom in a 100~bar hydrogen atmosphere) and thus a much larger fraction of the muons are able to reach the target compared to the case of a pure hydrogen gas. 

In the final step, the tightly bound $\mu d$ atom easily passes through the electron cloud of the target atoms as it behaves like a small neutral particle. Once close enough to the target nucleus, the muon is transferred for the last time and forms the desired muonic atom with the target nucleus $\mu  A$
\begin{equation}
\mu  d + A \rightarrow d +\mu  A^* .
\end{equation}
The excited muonic atom $\mu  A^*$ de-excites through emission of Auger electrons and X-rays to the ground state (compare Section \ref{sec_xrays}), which allows the measurement of the muonic X-ray spectrum.

\subsection{Setup}\label{sec_setup}
\begin{figure}
  \centering
\begin{subfigure}{.45\textwidth}
\centering
\includegraphics[width=0.87\textwidth]{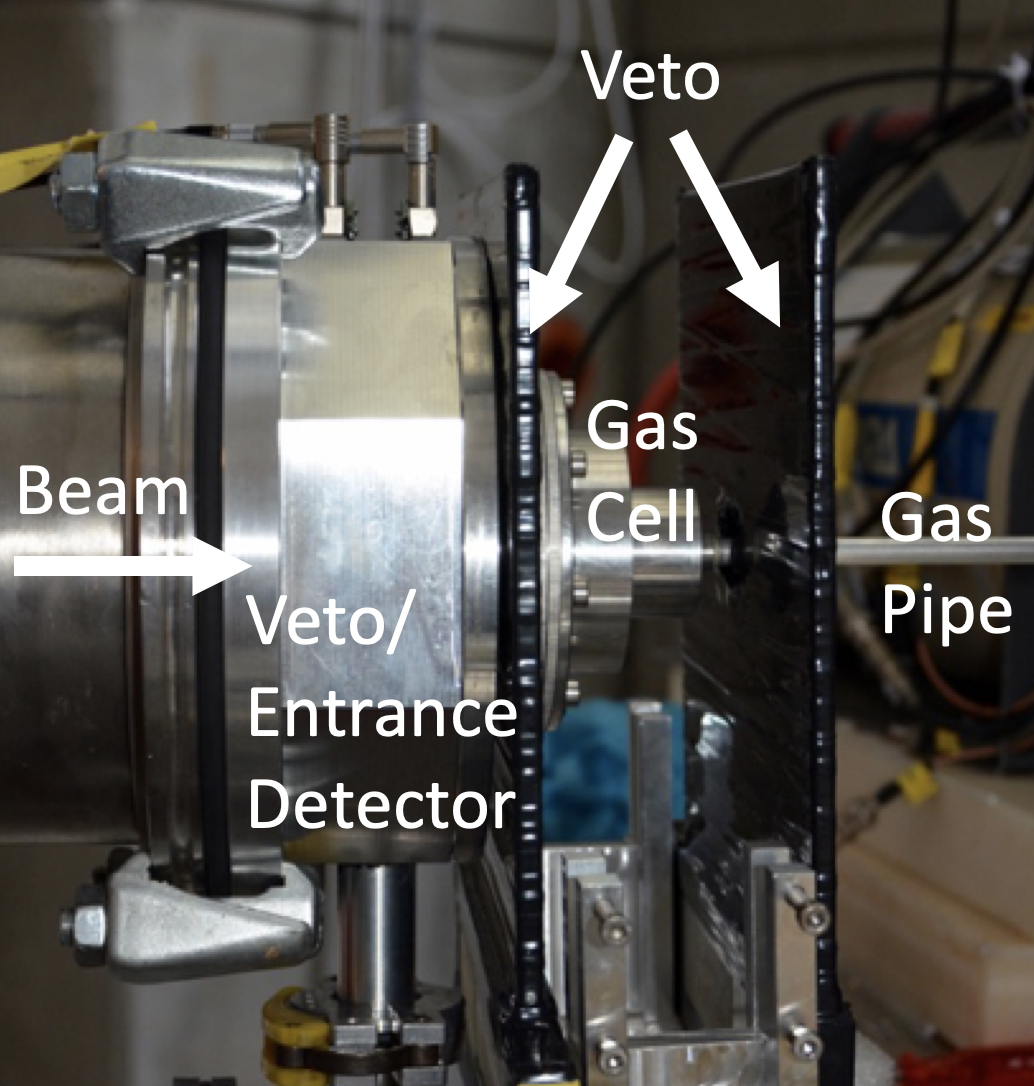}
\caption{Close-up of the gas cell together with the entrance and veto detectors.}
\label{Fig:CellSetup}
\end{subfigure}
\begin{subfigure}{.45\textwidth}
\centering
\includegraphics[width=0.97\textwidth]{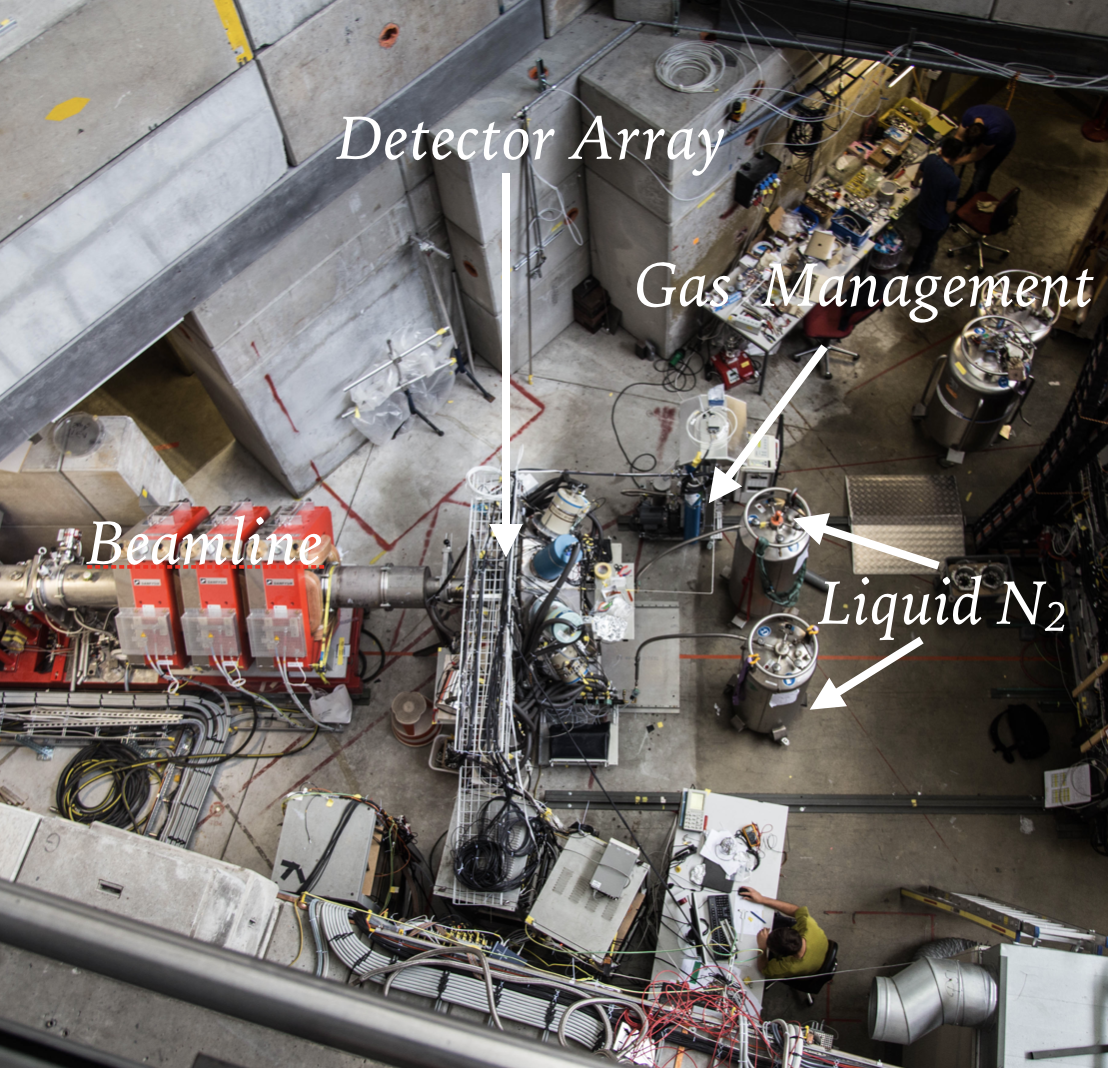}
\caption{Overview of the $\pi$E1 area showing the set up of 2017.}
\label{Fig:OverviewSetup}
\end{subfigure}
\caption{Experimental setup for the $\SI{5}{\mu g}$ gold target measurement.}
\end{figure}
A sketch of the experimental setup was already shown in Fig.~\ref{Fig:ExpSetup}. The following paragraphs will describe in more detail the different elements.

\begin{figure}
\centering
\includegraphics[width = 0.5\textwidth, angle=270]{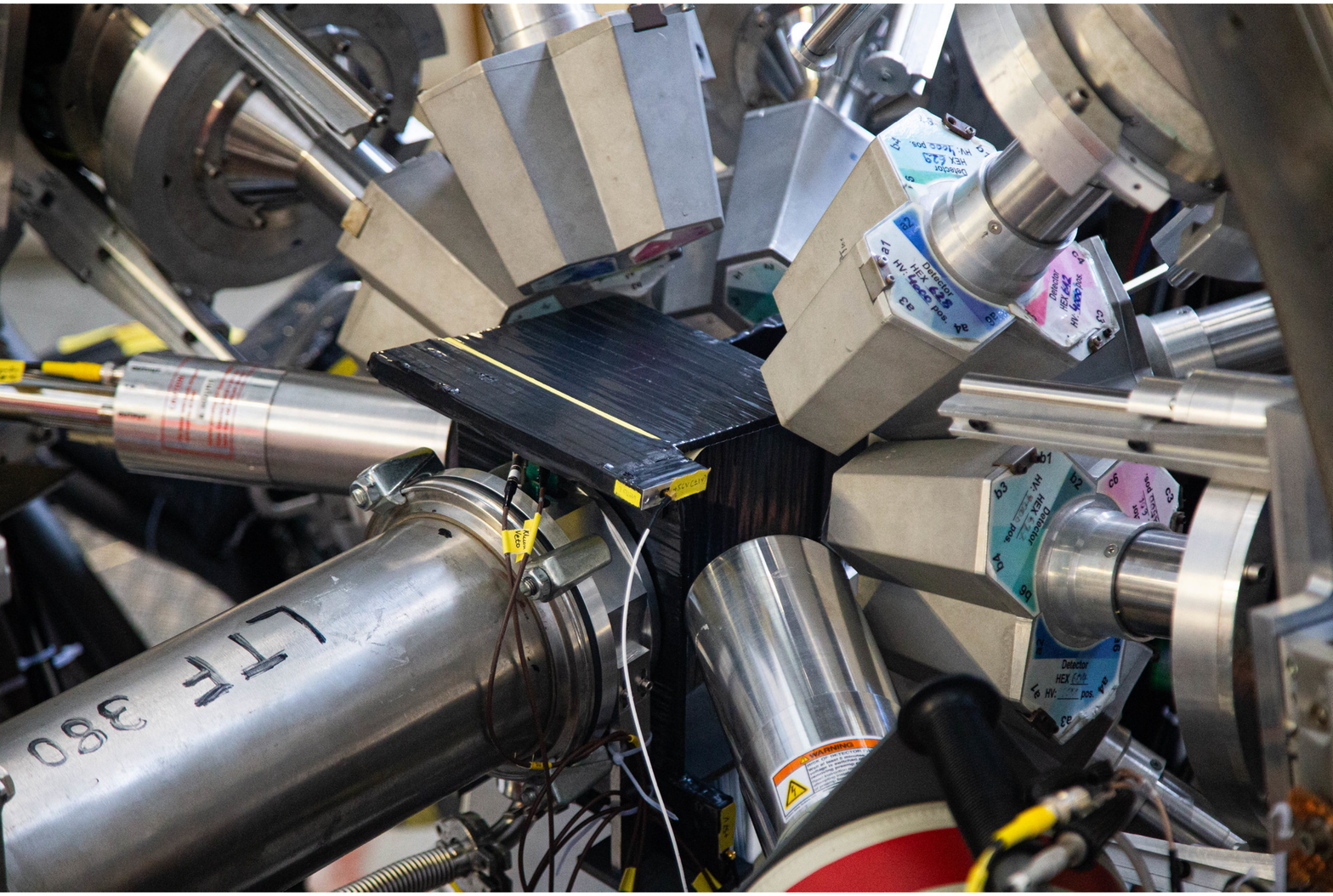}
\caption{For the measurement of the $^{226}$Ra and $^{248}$Cm targets the Miniball germanium detector array \cite{war13} was set up in the $\pi$E1 area.}
\label{Fig:Miniball}
\end{figure}

The experiment takes place in the $\pi$E1 area of the HIPA facility at PSI. The $\pi$E1 area offers a high purity negative muon beam with a muon rate of several $\SI{e+4}{} $  $\mu$/$s$ at our chosen momentum. An overview of the $\pi$E1 area including the muX experiment is shown in Fig.~\ref{Fig:OverviewSetup}. The beam enters the experimental setup from the left side. The red magnets are quadrupole magnets focussing the beam onto the gas cell. The gas cell shown in Fig.~\ref{Fig:CellSetup} is surrounded by an array of HPGe detectors. While in 2017 and 2018, most of the HPGe detectors were obtained from the French/UK germanium detector loan pool \cite{loan_pool}, in 2019 the Miniball germanium detector array \cite{war13} was used for the muonic atom X-ray measurements of $^{226}$Ra and $^{248}$Cm (compare Figure \ref{Fig:Miniball}). All HPGe detectors require constant cooling with liquid nitrogen. For this purpose, an automated cooling system for the detector array was installed.

Incoming muons are detected by a $\SI{20}{}\times \SI{20}{mm\tothe{2}}$ large and $\SI{100}{\mu m}$ thin plastic scintillator. In front of this entrance detector is a $\SI{5}{mm}$ thick plastic scintillator including a circular hole with a diameter of $\SI{15}{mm}$. This detector serves as a collimator and veto detector at the same time.  A layer of $^{208}$Pb is placed in front of this veto detector for the continuous measurement of lead muonic X-rays. Using the precisely measured muonic X-rays of $^{208}$Pb \cite{Ber88} the HPGe detectors can continuously be energy calibrated. As additional calibration means, a $^{60}$Co $\gamma$-ray source mounted next to the target, natural background lines like the $\SI{2614.5}{keV}$ $\gamma$-ray of the thorium series or accidental muonic lines from the target setup like $^{12}$C or $^{27}$Al are used.

The gas cell itself is a cylindrical aluminum vessel with a length of $\SI{2}{cm}$ and a diameter of $\SI{30}{mm}$ closed towards the muon beam by an entrance window made of a $\SI{0.6}{mm}$ thin carbon fibre sheet. It is reinforced by a $\SI{4}{mm}$ thick titanium support grid with an additional $\SI{2}{mm}$ thick carbon fiber grid on top. This configuration is robust enough to withstand the pressure of $\SI{100}{bar}$ inside the gas cell. The target is mounted on a plate at the back wall of the cell opposite to the entrance window. The wall of the cell behind the target also contains the inlet/outlet for the gas allowing to fill the hydrogen/deuterium gas mixture.

\subsection{Transfer optimization \label{sec_optimization}}
The gas cell is designed to stop muons in $100$ bar of the hydrogen-deuterium gas mixture at room temperature. This corresponds to $\sim$10\% of the density of liquid hydrogen.  The remaining parameters to optimize the process of transferring muons to a target mounted at the back wall of the gas cell are the stopping position of the muon beam and the amount of deuterium admixture to the hydrogen gas. 

Concerning the admixture of deuterium it is necessary to balance two opposing effects. Additional deuterium increases the transfer rate from muonic hydrogen to deuterium. However, the additional deuterium decreases the mean diffusion radius of the muonic deuterium atom due to the increased $\mu d$-D$_2$ scattering. In order to determine the best gas mixture for the experiment, the number of muonic X-rays per muon is measured for a given target for different gas mixtures. It was found that a deuterium admixture of 0.1 to 0.25\% is optimal.

The muon beam with a momentum of $p\simeq\SI{28}{MeV/c}$ results in a stopping distribution with a full width at half maximum of approximately $\SI{7}{mm}$ in the gas mixture. The stopping position, i.e. the center of the stopping distribution, can easily be tuned by changing the beam momentum. Muons with a \SI{1}{MeV/c} lower momentum stop too close to the entrance window decreasing the probability of reaching the target. In contrast, muons with a \SI{1}{MeV/c} higher momentum are not stopped in the gas but in the backing plate of the target resulting in the direct atomic capture in the backing plate material. As before, the number of observed muonic X-rays is measured for a given target material and the beam momentum carefully tuned to maximize the yield of these X-rays.

\subsection{Transfer Efficiency}
In 2017 the muX collaboration demonstrated for the first time that the newly developed procedure transfers muons to a gold target with a mass of only $\SI{5}{\mu g}$ \cite{ska19}. Due to the expected small transfer rate to the  $\SI{5}{\mu g}$ target, the test started with the optimizations described in Section~\ref{sec_optimization} for a $\SI{50}{nm}$ thick gold target evaporated onto a copper backing foil with a surface of $\sim\SI{4.9}{cm^2}$ corresponding to a target mass of almost 500~$\mu$g. After reaching the optimal settings for this target, the target was replaced by a $\SI{10}{nm}$ thick gold target and the optimization procedure was repeated. The next optimization step involved a $\SI{3}{nm}$ thick target with the same surface area. The successive reduction of the target mass ensured a proper evaluation of the required measurement time and experimental settings for the $\SI{5}{\mu g}$ target.

\begin{figure}
\centering
\begin{subfigure}{.48\textwidth}
  \centering
  \includegraphics[width=1\linewidth]{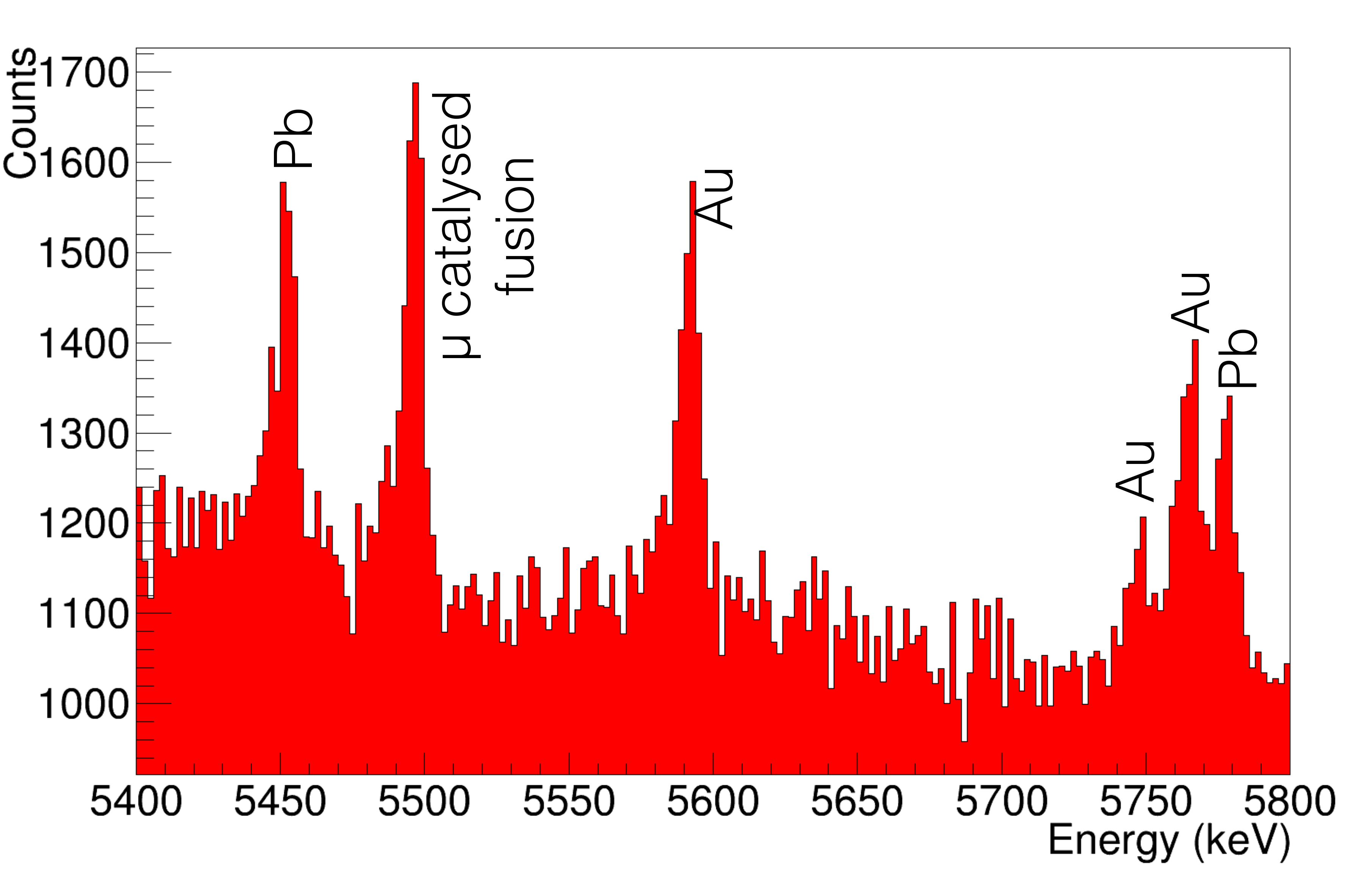} 
  \caption{}
  \label{fig:Spectrum1D}
\end{subfigure}%
\begin{subfigure}{.48\textwidth}
  \centering
  \includegraphics[width=1\linewidth]{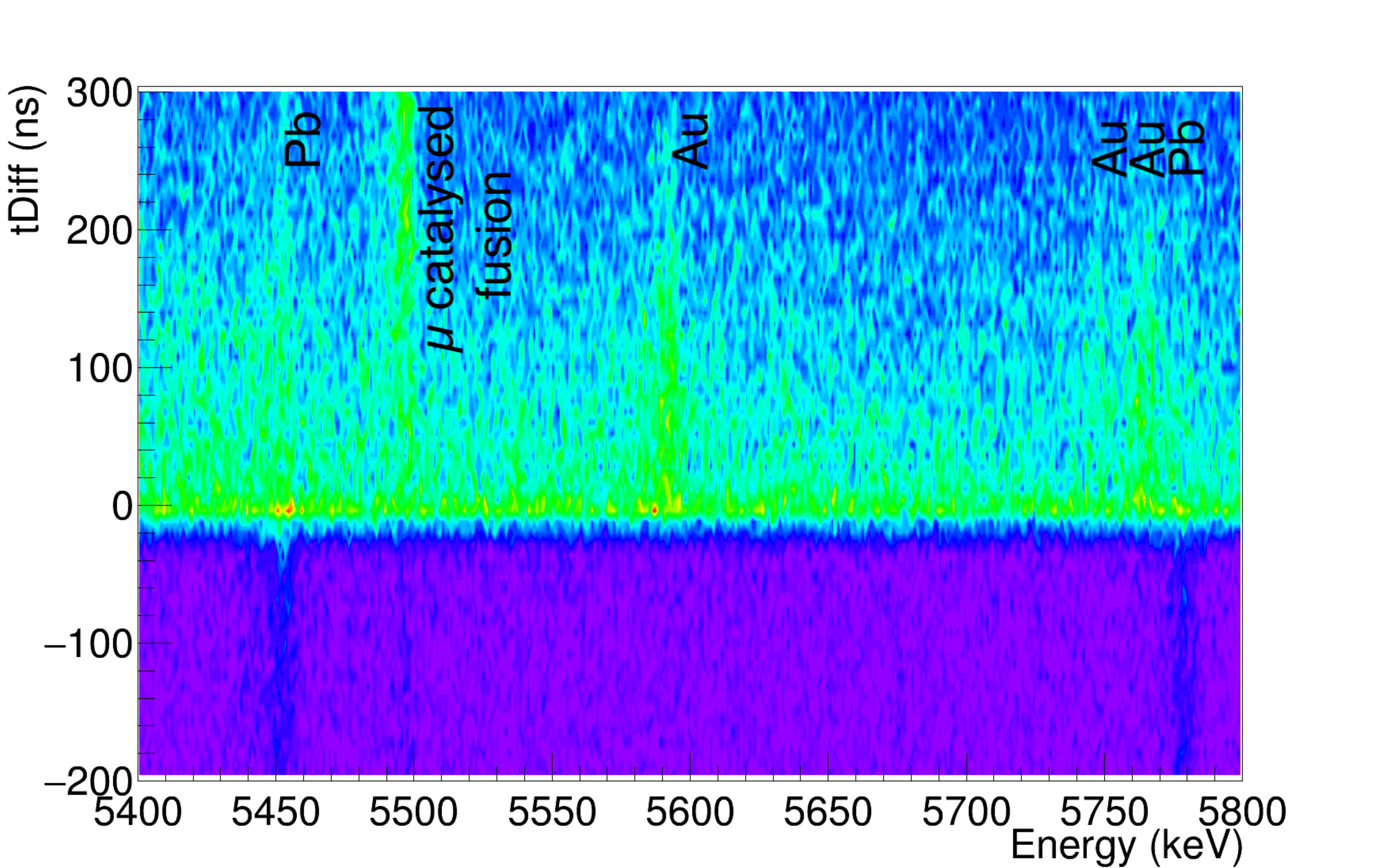} 
  \caption{}
  \label{fig:Spectrum2D}
\end{subfigure}
\caption{(Left) Gold muonic X-rays as measured from a 5~$\mu$g gold target over 18.5~h. The $2p-1s$ transitions in gold are the lines at $\SI{5.59}{MeV}$, $\SI{5.75}{MeV}$ and $\SI{5.77}{MeV}$. The calibration X-rays of $^{208}$Pb are measured at $\SI{5.45}{MeV}$ and $\SI{5.78}{MeV}$. Moreover, an X-ray from muon catalyzed fusion is observed at $\SI{5.5}{MeV}$. (Right) For each event in the HPGe detector the time difference (tDiff) between the signal in the HPGe and in the entrance detector was determined. This additional information is presented in the 2D spectrum and allows to differentiate uncorrelated X-rays such as from $^{208}$Pb (constant in time) from prompt X-rays (centered around tDiff=0) and X-rays stemming from muon transfers (starting at tDiff=0 and extending for a few hundred ns). }
\end{figure}

The final gold target featured the mentioned mass of $\SI{5}{\mu g}$ corresponding to a $\SI{3}{nm}$ thick coating on a $\SI{1}{cm^2}$ area. After a total measurement time of $\SI{18.5}{h}$, the HPGe detectors revealed the muonic atom energy spectrum of the gold $2p-1s$ region shown in Fig.~\ref{fig:Spectrum1D}. The $2p-1s$ lines of muonic  $^{208}$Pb used for calibration purposes are in the same energy region. In addition, a line stemming from muon catalysed fusion is observed. The muon catalysed fusion is a clear indicator for a successful muon transfer from muonic hydrogen to deuterons. The corresponding two-dimensional spectrum showing in addition time information is shown in Fig.~\ref{fig:Spectrum2D}. The time difference tDiff for the HPGe detector signals is calculated with respect to the correspondingly detected muon in the entrance detector. Background signals, such as the calibration X-rays, and the desired muonic X-rays can in principle be differentiated by carefully analyzing  this 2D spectrum. 

The spectrum was measured with a momentum of $p = \SI{27.75}{MeV/c}$ and a deuterium admixture of $\SI{0.25}{\%}$ in a  \SI{100}{bar}  hydrogen/deuterium gas mixture at room temperature. The test showed that about $\SI{1}{\%}$ of the stopped muons in the gas cell were transferred to the $\SI{5}{\mu g}$ gold target \cite{ska19}.

\section{Conclusions} \label{sec_conclusions}
Muonic atoms can be used to study nuclear properties with high sensitivity due to the close orbits of the muon around the nucleus. By measuring the energy of the characteristic X-rays emitted by the captured muon on its cascade down to the ground state -- a technique termed muonic atom spectroscopy -- the full muonic energy level scheme can be readily reconstructed. Two nuclear properties of interest can be extracted in that way from the measured X-ray spectrum: the absolute nuclear charge radius and the quadrupole moment of the nucleus. While the nuclear charge radius leads to overall shifts of the energy levels, the quadrupole moment leads to a hyperfine structure of the muonic transitions.

Almost all stable elements have been subject of muonic atom spectroscopy campaigns with corresponding extractions of their charge radii and quadrupole moments (where applicable). However, no elements that can be handled only in small quantities -- either due to their scarcity or radioactivity -- could be probed so far. The muX experiment has developed a method based on muon transfer reactions inside a high pressure hydrogen/deuterium gas cell to probe targets available only in microgram quantities. A first proof-of-principle with a target of 5~$\mu$g of gold was very successful and a nice muonic energy spectrum could be recorded. In 2019, measurements with the radioactive targets $^{248}$Cm and $^{226}$Ra were performed, the analysis of which is currently ongoing.

\section{Acknowledgements} \label{sec_ack}
We are very grateful to all the members of the muX collaboration belonging to the following institutions: Polish Academy of Sciences (Poland), Paul Scherrer Institut (Switzerland), ETH Z\"urich (Switzerland), Johannes Gutenberg University Mainz (Germany), KU Leuven (Belgium), GSI Helmholtzzentrum f\"ur Schwerionenforschung Darmstadt (Germany), Helmholtz Institute Mainz (Germany), Universit\"at zu K\"oln (Germany), LKB Paris (France), University of Groningen (The Netherlands), University of Pisa and INFN, Pisa (Italy), University of Victoria (Canada), Perimeter Institute (Canada), CSNSM, Universit\'e Paris Saclay (France). The experiments were performed at the $\pi$E1 beamline of PSI. We would like to thank the accelerator and support groups for the excellent conditions. Technical support by F. Barchetti, F. Burri, M. Meier and A. Stoykov from PSI and B. Zehr from the IPP workshop at ETH Z\"urich is gratefully acknowledged.  This work was supported by the Swiss National Science Foundation through project grant No. 200021\_165569 and the German BMBF under contract 05P18PKCIA.

%
%
%

%
%
%

\end{document}